\documentclass[twocolumn,superscriptaddress]{revtex4-2}
\usepackage{xcolor}
\usepackage{amsmath, amsfonts, amssymb, graphicx, color}

\usepackage{url}

\newcommand{\lastequal}{Corresponding authors. These authors contributed equally.}

\begin{document}

\newcommand{\deftitle}{{Computational detection of antigen specific B cell receptors following immunization}}

\title{\deftitle}

\author{Maria Francesca Abbate}
\affiliation{Laboratoire de physique de l'\'Ecole normale sup\'erieure,
  CNRS, PSL University, Sorbonne Universit\'e, and Universit\'e de
  Paris, 75005 Paris, France}
\affiliation{Large Molecule Research, Sanofi, Vitry-sur-Seine, France}
\author{Thomas Dupic}
\affiliation{Department of Organismic and Evolutionary Biology, Harvard University, Cambridge, United States}
\author{Emmanuelle Vigne}
\affiliation{Large Molecule Research, Sanofi, Vitry-sur-Seine, France}
\author{Melody A. Shahsavarian}
\affiliation{Large Molecule Research, Sanofi, Vitry-sur-Seine, France}
\author{Aleksandra M. Walczak}
\thanks{\lastequal}
\affiliation{Laboratoire de physique de l'\'Ecole normale sup\'erieure,
  CNRS, PSL University, Sorbonne Universit\'e, and Universit\'e de
  Paris, 75005 Paris, France}
\author{Thierry Mora}
\thanks{\lastequal}
\affiliation{Laboratoire de physique de l'\'Ecole normale sup\'erieure,
  CNRS, PSL University, Sorbonne Universit\'e, and Universit\'e de
  Paris, 75005 Paris, France}

\begin{abstract}

B cell receptors (BCRs) play a crucial role in recognizing and fighting foreign antigens. High-throughput sequencing enables in-depth sampling of the BCRs repertoire after immunization. However, only a minor fraction of BCRs actively participate in any given infection. To what extent can we accurately identify antigen-specific sequences directly from BCRs repertoires? 
We present a computational method grounded on sequence similarity, aimed at identifying statistically significant responsive BCRs. This method leverages well-known characteristics of affinity maturation and expected diversity. We validate its effectiveness using longitudinally sampled human immune repertoire data following influenza vaccination and Sars-CoV-2 infections. We show that different lineages converge to the same responding CDR3, demonstrating convergent selection within an individual. The outcomes of this method hold promise for application in vaccine development, personalized medicine, and antibody-derived therapeutics.

\end{abstract}

\maketitle

\section{Introduction}

\begin{figure} 
    \centering
    \includegraphics[width=\linewidth]{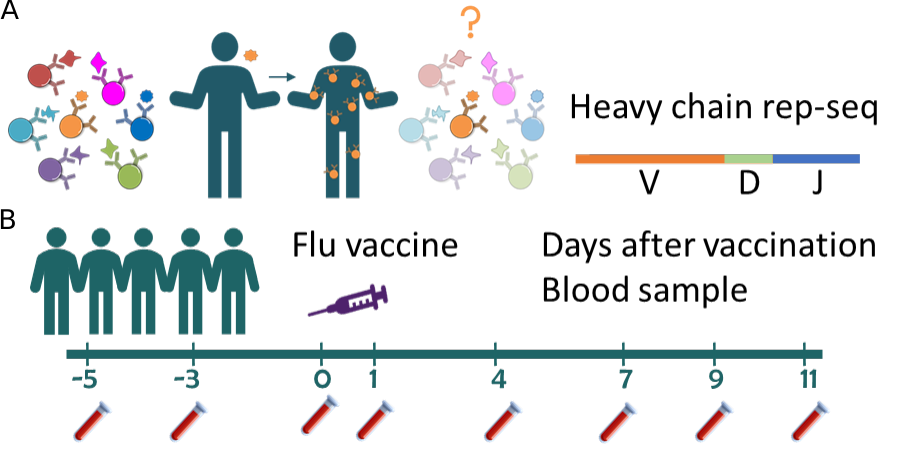}
    \caption{{\bf Identifying responding antibodies from repertoires. A.} B cell repertoires exploit a diverse set of  antigen receptors (antibodies) with different antigen specificities. Upon an immune challenge, antigen-specific B cells proliferate and mutate. The question addressed here is how to identify these responding clones from repertoire data.
      {\bf B.} We exploit bulk BCRs immune repertoire sequencing data~\cite{Horns1} that covers the V, D and J segments of the BCRs heavy chain to  detect influenza--responding B cells without knowing the epitope, using the repertoire sampled at a single timepoint. Five healthy humans were vaccinated in late spring of 2012 with the 2011–2012 trivalent seasonal flu vaccine. Blood samples were collected before (days -5, -3, and 0) and after (1, 4, 7, 9 and 11) vaccine administration.}
    \label{Cartoon_1}
  \end{figure}

B cells in the adaptive immune system express receptors on their surface that bind parts of proteins, called antigens, to iniate an immune response  (Fig. \ref{Cartoon_1}).
Identifying B cell receptors (BCRs) that respond to specific antigens is an important goal for describing the dynamics of B cell immunity following vaccination, with potential applications in vaccine development \cite{Fink,Wang2,Galson,Wang3,Kreer,Kreer3}, personalized medicine \cite{Hoehn_cut,Wu,Ota}, and antibody-derived therapeutics \cite{Kreer2,Gieselmann,Agrafiotis,Pedrioli,Robinson}.

The initial diversity of immune receptor repertoires is generated through the random assembly of genomic templates complemented by deletions and insertions at the gene junctions. This initial diversity is further enhanced by affinity maturation.
Upon antigenic stimulation, B cells that recognize an antigen migrate to germinal centers, where they acquire somatic hypermutations in their antigen receptor, and undergo selection for antigen binding. This process ultimately results in B cells that better recognize the antigen. Since B cells are released to the periphery throughout, affinity maturation produces lineages of related B cells that can be identified by sequence similarity of their antigen receptor \cite{Natanael,Matsen,nouri2020somatic}.  As a result, cells carrying different but similar sequences are involved in neutralizing the same antigen with different potencies. In addition, several distinct founder B cells specific to the same antigen can seed distinct but related lineages~\cite{Victora}. It is still unclear how affinity maturation further diversifies and focuses the responding repertoire.

Recent advances in high-throughput sequencing technologies make it possible to directly profile the immune repertoire by sequencing the B cell DNA or mRNA taken from blood or tissue samples (RepSeq)~\cite{weinstein2009high,baum2012wrestling,robins2013immunosequencing,
vollmers2013genetic,Georgiou,Briney,kreer2022probabilities}. 
The experiments provide a list of unique sequences with their relative abundances.
However, exploiting repertoire information for decoding the immune response is hindered by both our inability to reliably predict the specificity of a given BCR to a given antigen, and the fact that only a small fraction of the repertoire is involved in any infection (Fig. \ref{Cartoon_1}A). Existing methods for identifying responding BCRs combine traditional sorting assays and sequencing \cite{kreer2020exploiting,babcook1996novel,harding2010immunogenicity,
Gieselmann,Rapid} with computational analyses \cite{Sorting}. Approaches based on machine learning \cite{Greiff,Akbar,Kunik,Jespersen,Liberis,Mason},  network analysis \cite{Miho,Hoehn,Chang}, publicness across multiple individuals~\cite{Osbourn,
Montague2021,ruiz2023modeling}, or using structural information \cite{Wong,Kova,Structure,raybould2019antibody,robinson2021epitope}, have also been proposed to focus on the disease-specific sub-repertoire.
Longitudinal sampling of repertoires after a strong antigenic stimulation, such as a vaccine or disease~\cite{Horns1, Galson,laserson2014high}
is an agnostic way to study the response that does not require prior knowledge about the identity of the triggering antigens, and can identify a polyclonal response against multiple epitopes. Such approaches have also been successfully applied to T-cell repertoires \cite{pogorelyy2018precise,minervina2021longitudinal,puelma2020inferring,minervina2020primary} where the identity of epitopes can sometimes be reverse-mapped using specificity databases \cite{pogorelyy2022resolving}.
These methods directly rely on clonotype abundance measured in the repertoire, offering alternatives to traditional screening assays. However they require to obtain blood samples from individuals prior to the immune challenge, which is not always practical.

We propose a computational approach for identifying clusters of expanded BCRs from the repertoire measured at a single time point, by combining information about sequence similarity and convergent selection. Such convergence has previously been exploited to identify responding clonotypes in the context of T cell repertoires, and is the basis of the ALICE software tool~\cite{Alice}. However, affinity maturation and lack of HLA restriction make the problem very different for B cells. 
We illustrate our method on data from recent studies that track the unbiased B cells responses of 5 healthy individuals after influenza immunization with the 2011–2012 trivalent seasonal flu vaccine in late spring of 2012 ~\cite{Horns1}  (Fig. \ref{Cartoon_1}B), and 18 recent COVID-19 patients at the peak of  infection ~\cite{Galson}. This allows us to study the multiplicity, diversity and convergent selection of distinct lineages towards viral antigens with unprecedented details.

\section{Results}

\subsection*{Computational identification of responding clones from a single timepoint}

\newcommand{\mn}{STAR}

\begin{figure*} 
    \centering
    \includegraphics[width=\textwidth]{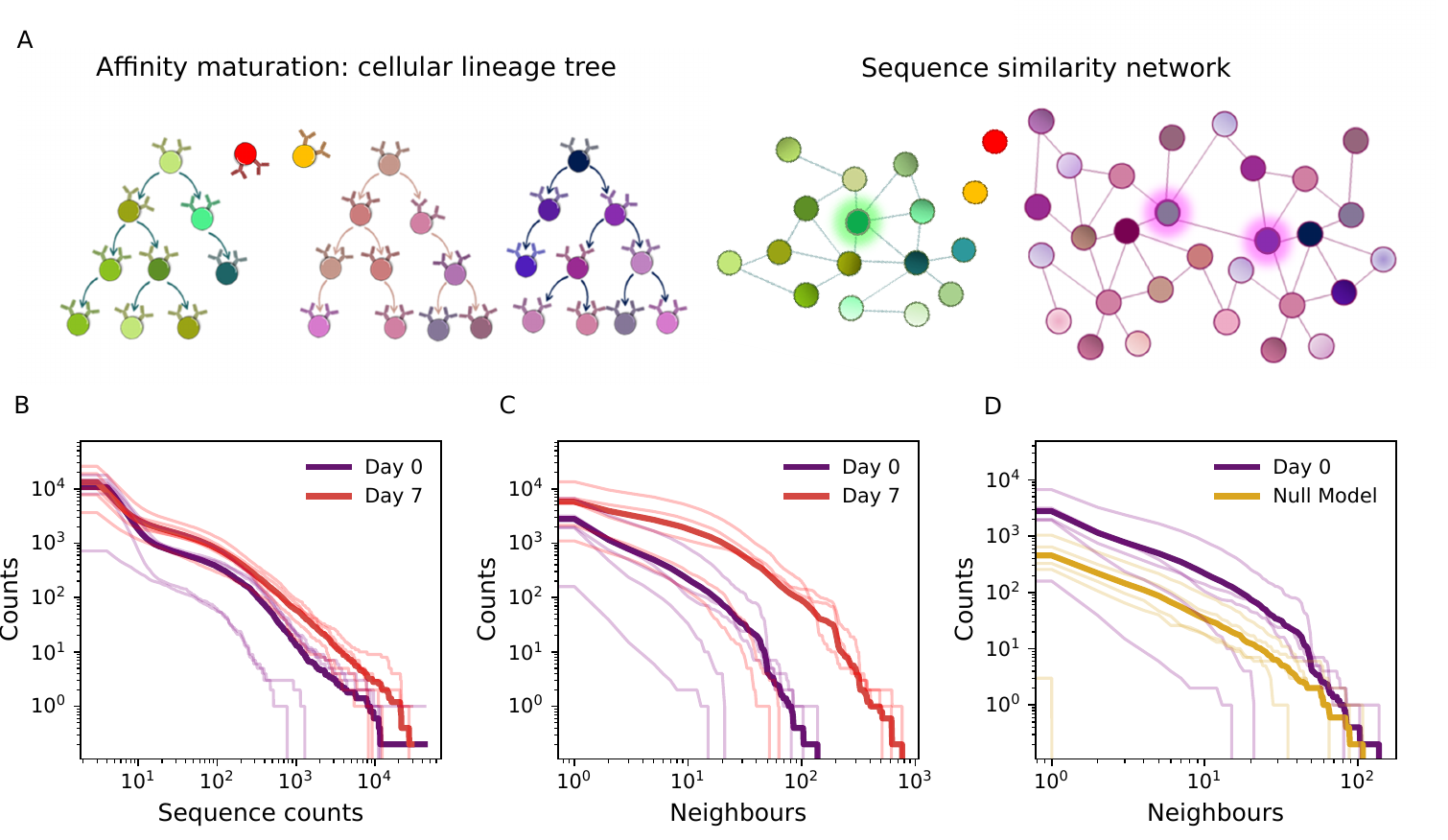}
    \caption{
\textbf{Similarity network analysis of antibody clonotypes. A.} During affinity maturation, distinct naive B cells proliferate and mutate upon recognition of the antigens, giving rise to distinct cell lineages (left). At the sequence level (right), we construct a graph where each node is an IgH nucleotide sequence, and edges connect sequences that differ by at most one amino acid in their CDR3. This may lead to distinct lineages being merged into the same functional cluster (e.g. the pink cluster). The idea of {\mn} is to identify sequences with high connectivity in the graph (highlighted), which indicates either convergent selection or belonging to a large lineage, or both. \textbf{B.} Distribution of the CDR3 amino acid sequence count for all subjects (background lines) and their average (thick lines), at days 0 and 7. The two distributions are similar \textbf{C.} The distribution of the number of amino acid neighbours shows a marked difference between days 0 and 7 (same color convention as B). \textbf{D.} The distribution of neighbors at day 0 is well described by a computational model of random repertoire generation \cite{OLGA} (see main text).}
    \label{Fig_distribution}
\end{figure*}

When a B cell is involved in an immune response, its BCR undergoes proliferation and mutations. This process yields many copies of the original BCR, while also producing mutated receptors with similar sequence and antigen specificity as the ancestral BCR. At the repertoire level, we expect two main effects: elevated frequencies of responding clonotypes, and the formation of extensive clusters of similar sequences. These clusters may arise both from the expansion of a single BCR lineage, and from the convergent selection of multiple lineages with distinct progenitors (Fig. \ref{Fig_distribution}A).

We first examined the profile of clonotype frequencies in the BCRs heavy chain (IgH) repertoires of five healthy individuals who had received the 2011–2012 trivalent seasonal flu vaccine (see Methods and Table S1 for details), both before the vaccine (day 0), and at the peak of the response (day 7) \cite{Horns1}. A clonotype is defined by the unique amino acid sequence coding for the Complementarity Determining Region 3 (CDR3) of the heavy chain. The comparison of the distributions of clonotype abundances shows very little change between before vaccination and at the peak of the response (Fig. \ref{Fig_distribution}B), challenging the expectation that the post-vaccination repertoire should be dominated by a few large responding clonotypes.

We next consider the \textit{number of neighbors} in amino acid space as a measure of sequence similarity. For each CDR3 amino acid sequence found in the repertoire, we count the number of distinct IgH nucleotide sequences whose CDR3 differ by exactly one amino acid. Accounting for the mulitiplicity of nucleotide sequences helps capture the diversity of convergent synonymous variants in the response. In contrast to frequencies, the distribution of the number of neighbors does change significantly (Fig. \ref{Fig_distribution}C) from day 0 to day 7, in agreement with the expectation that mutations and convergent selection can create clusters of related sequences.

We use this observation to introduce two approaches to identify antigen-specific B cell receptors. The approaches extend previous ideas proposed for T cells~\cite{Alice} to the context of affinity maturation. The first, called fast-{\mn}, (fast Single Timepoint Antibody Recognition) prioritises computational speed and performs efficient thresholding to output sequences with a high confidence level as responders. 
The second, less specific method, called full-{\mn}, assigns a score to each sequence based on a probabilistic approach.

Specifically, fast-STAR calls an IgH CDR3 amino acid sequence a responding clonotype (a `hit') if its number of neighbors, normalized by the total number of unique nucleotide sequences, is higher than a certain threshold. This threshold is set to $9.6\cdot 10^{-4}$ to obtain a false-discovery rate (FDR) of $\approx 5\%$, as estimated by taking the ratio of the number of hits at days -3 and 0 (only false positives) with the number of hits at day 7 (true positives and false positives), averaged over all 5 subjects.
Sequences above the threshold are then grouped by single-linkage clustering, where two amino acid sequences are linked if they have Levenshtein distance 1 or lower. Clusters with fewer than 10 sequences are filtered out to mitigate the effect of potential sequencing errors. As a result, fast-STAR only keeps a small number of hits.

This method results in very high specificity but low sensitivity. In addition, it does not exploit the knowledge that certain sequences are expected to have more neighbors than other due to biases in the generation probability.
Full-{\mn} overcomes these limitations by setting a personalized sequence-dependent threshold for the number of neighbors based on the expectation computed from the previously proposed OLGA method~\cite{OLGA}, which estimates the probability of observing any given sequence in a random repertoire (see Methods). We can use two methods to estimate the number of neighbors using OLGA: either exactly, by summing the probabilities of generation of each possible neighbor for each observed sequence, or approximately, by exploiting the fact that neighboring sequences have similar probabilities of generation (Fig.~\ref{Pgen_Neigh}) to reduce computational time (see Methods and Fig.~\ref{Lambda_neigh}).
OLGA's prediction for the distribution of neighbors using the second approximate method agrees well with that measured at day 0 (Fig. \ref{Fig_distribution}D), and so subsequently we will only use that method.
Responding sequences are identified as having more neighbors than expected by the model, with significance computed using the Poisson distribution. The threshold on the resulting p-value is controled for multiple testing by the Benjamini-Hochberg procedure by setting a false discovery rate of $0.5\%$.

\begin{figure*} 
    \centering
    \includegraphics[scale=0.7]{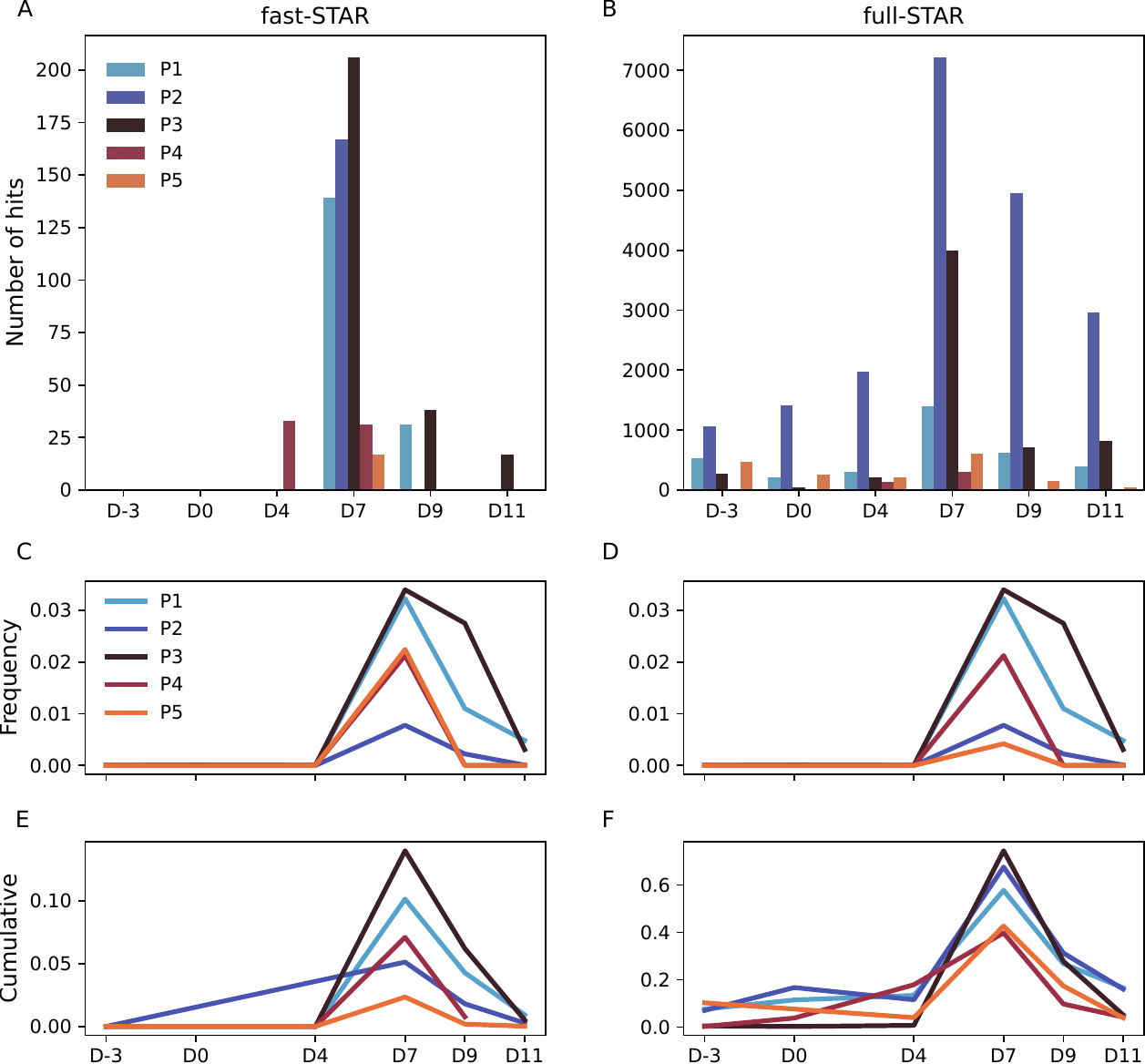}
    \caption{
        {\bf Results and validation.} \textbf{A.-B.} Number of putative responding sequences identified by ({\bf A}) fast-{\mn} and ({\bf B}) full-{\mn} for each day and each subject. We observe very few hits on the days before vaccination and a large peak on day 7. \textbf{C.-D.} Frequency time trace of the top-scoring sequence for each subject found by (\textbf{C}) fast-{\mn} (with largest number of neighbors) and (\textbf{D}) full-{\mn} (with lowest p-value). Note that the best-scoring clonotype is the same for the two methods for all subjects except subject 5.
      \textbf{E.-F.} Sum of the frequencies of all responding clonotypes according to (\textbf{E}) fast-{\mn} and (\textbf{F}) full-{\mn}.}
    \label{Fig_validation}
\end{figure*}

\subsection*{Clones identified by {\mn} recapitulate the immune response dynamics}

We applied both fast-{\mn} and full-{\mn} to the IgH repertoires of all 5 subjects at each timepoint {(SI Table S1)}. Although neither pipeline used any information about the time course of clonal abundances, they could both detect a marked increase of the number of hits following vaccination, with a peak on day 7 and rapid decay after that (Fig. \ref{Fig_validation} A-B).
This response peak is consistent with previous observations based on longidutinal analysis \cite{Horns1}, and is characteristic of a memory recall response following vaccination. The fast-{\mn} method, which is more conservative and robust, even finds no hits at all for all the subjects prior to day 7. This suggests that the pipeline specifically identifies responding clonotypes.
The most common isotype found in the repertoires is IgM, except at day 7 where it is still substantially represented (Fig.~\ref{IgClass}A). However, almost all clonotypes identified by fast-{\mn} in subject 1 were IgG  (Fig.~\ref{IgClass}B), while full-{\mn} hits were IgA or IgG (Fig.~\ref{IgClass}C), with most IgA found before the response peak. This serves as further validation that the clonotypes identified by {\mn} are involved in the memory recall response, and suggests that those identified by full-{\mn} also include previously expanded clones associated to distinct immune challenges.

Clonotypes identified by {\mn} on day 7 can be used to retrospectively study the dynamics of the response. Examining the frequency time courses of single clonotypes identified as responding shows a consistent pattern across all subjects, with no detected presence before day 7, and a sharp peak at day 7 followed by rapid decay  (Fig.~\ref{Fig_validation}~C-D). This again validates the approach, as neither pipeline used frequency information. To go beyond single clonotypes, we agregated the frequencies of all clonotypes identified as responsive, and plotted their cumulative frequency as a function of time for both pipelines (Fig.~\ref{Fig_validation}E-F).
These time traces show again a marked peak on day 7 for both pipelines. The less conservative full-{\mn} captures a larger fraction of the responding repertoire, identifying as much as 75\% of the repertoire being involved in the response at its peak.

\begin{figure*} 
    \centering
    \includegraphics[width=.8\linewidth]{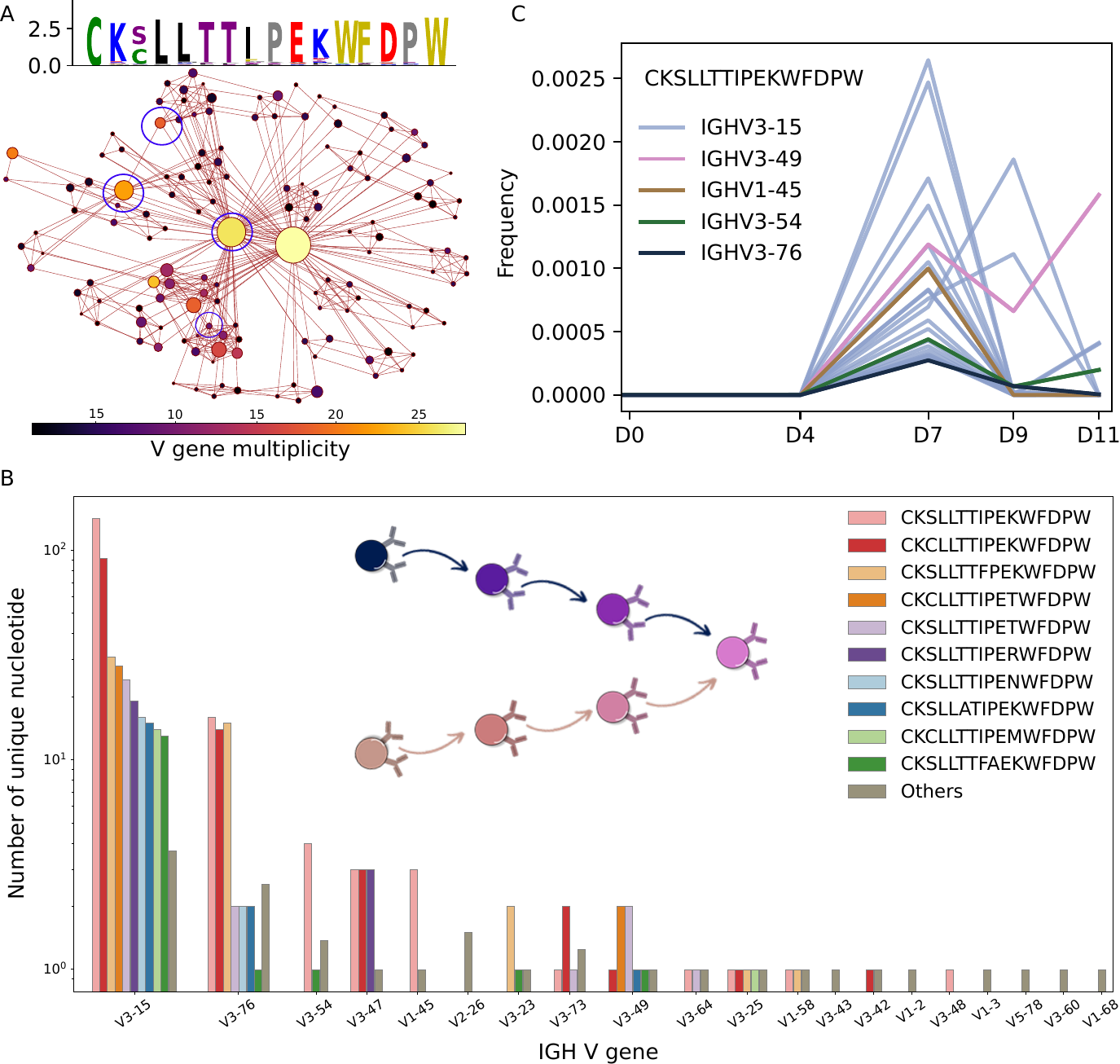}
    \caption{{\bf Convergent antibody response towards a conserved IgH CDR3 motif}.
      \textbf{A.} Graph structure of fast-STAR hits found in subject 1. Each node is an amino acid IgH CDR3 sequence. Node size is proportional to the number of distinct IgH nucleotide sequences with that CDR3, and color indicates the number of distinct V genes used. An edge is drawn between two CDR3 if they differ by one amino acid. The blue circle indicates the four amino acid CDR3s identified in the experimental testing as responding. Top: sequence logo of the sequences of the cluster. Height corresponds to the entropy of the amino acid choice at each site, and relative letter size to amino acid frequencies. \textbf{B.} Number of distinct nucleotide sequences with the same amino acid CDR3, grouped by V gene usage. Each CDR3 sequence can be formed using distinct V genes, and within each V gene group, up to hundreds of nucleotide variants.
      \textbf{C.} Frequency time course of the most abundant nucleotide sequences associated with the amino acid CDR3 sequence CKSLLTTIPEKWFDPW. Sequences with different V genes are shown in different colors. Each sequence corresponds to a distinct B cell clone that expanded independently at day 7.}
    \label{Conv_selection}
\end{figure*}

In addition, we directly validated the ability of some of our hits to bind the virus. In Ref.~\cite{Horns2}, 21 antibodies belonging to 5 distinct lineages found to be expanded in subject 1, and separately single-cell sequenced, were tested for affinity against the vaccine as well as various viral strains using ELISA assays. Among those 5, 2 lineages contained vaccine-binding antibodies, only one of which, called L1 and containing 5 antibodies, was also present in the bulk longitudinal data analyzed in the current paper. These 5 antibodies use 4 distinct CDR3s. All of them belonged to the cluster of hits identified by STAR in subject 1 (Fig.~\ref{Conv_selection}A), providing direct evidence that this group of antibodies specifically target the viral proteins. 

As a final validation test, we compared our putative influenza-responding clonotypes to those reported in~\cite{Kleinstein}, independently obtained from subjects vaccinated with the inactivated influenza vaccine. In \cite{Kleinstein} the authors computationally identified responsive sequences as those that significantly expanded between a pre- and post-vaccination timepoint. We evaluated the overlap between the IgH amino acid CDR3 sequence of their 1513 vaccine-responding candidate antibodies, and our STAR hits. While we found a small overlap of 2 sequences with our fast-STAR hits, and 9 with our full-STAR hits, these numbers are much larger than expected by chance (as evaluated by computing the overlap with a control dataset of healthy individuals ~\cite{Briney}, after normalizing for dataset size), $2\cdot 10^{-4}$ and $0.03$ respectively ($p=3\cdot 10^{-8}$ and $p=10^{-19}$, Poisson test).

\subsection*{Convergent selection}
The sequences identified by fast-STAR can be organized clusters of closely related sequences (Fig.~\ref{clusterlogo}). Three our of five subjects (1, 2, and 5) have a single cluster, suggesting an immunodominant response against a single epitope, while the other two (3 and 4) have four and three clusters respectively, suggesting a polyclonal response.
To better understand the structure of the repertoire response in sequence space, we analyzed in more detail the sequence structure of fast-STAR hits found in subject 1. These hits formed a single cluster of densely connected and highly conserved sequences (Fig. \ref{Conv_selection}A and weblogo therein).

We wondered whether this diversity of similar CDR3 amino sequences arose from a single lineage, or from distinct selection events originating from different naive sequences.
To explore this question, we considered the V genes of these sequences (given by MiXCR \cite{MIXCR}).
Multiple responding CDR3s were associated with different V genes, suggesting independent expansion and evolution of distinct B cell clones, with up to 23 different V genes associated with the same CDR3 (Fig. \ref{Conv_selection}B). This diversity of V gene assignment was not a spurious consequence of misassigment due to hypermutations, since the observed hypermutation rate (3 base pairs on average over the sequenced V region) was small compared to the pairwise distance between nearby V genes over the same region (at least 6 nucleotide differences, see Fig. \ref{Vgene}). 

To test this hypothesis, we examined the frequency dynamics of individual IgH nucleotide sequence variants coding for the top-scoring amino acid CDR3 of subject 1 (CKSLLTTIPEKWFDPW) (Fig.~\ref{Conv_selection}C). For illustration purposes, we took the most frequent nucleotide sequences (frequency $\geq 2.5\cdot 10^{-4}$), and color coded time traces by the germline V gene the sequences use. All traces show a clear and independent expansion between days 4 and 7, supporting the notion that multiple lineages independently converged towards a shared functional outcome. The fact that multiple distinct V genes are represented excludes the possibility of an artifact due to sequencing errors or hypermutations. This observation provides direct evidence for antigen-specific convergent selection of multiple lineages, and introduces potentially a novel feature for investigating and exploiting antigen-specific receptors.

\begin{figure*}[!th] 
    \includegraphics[width=\linewidth]{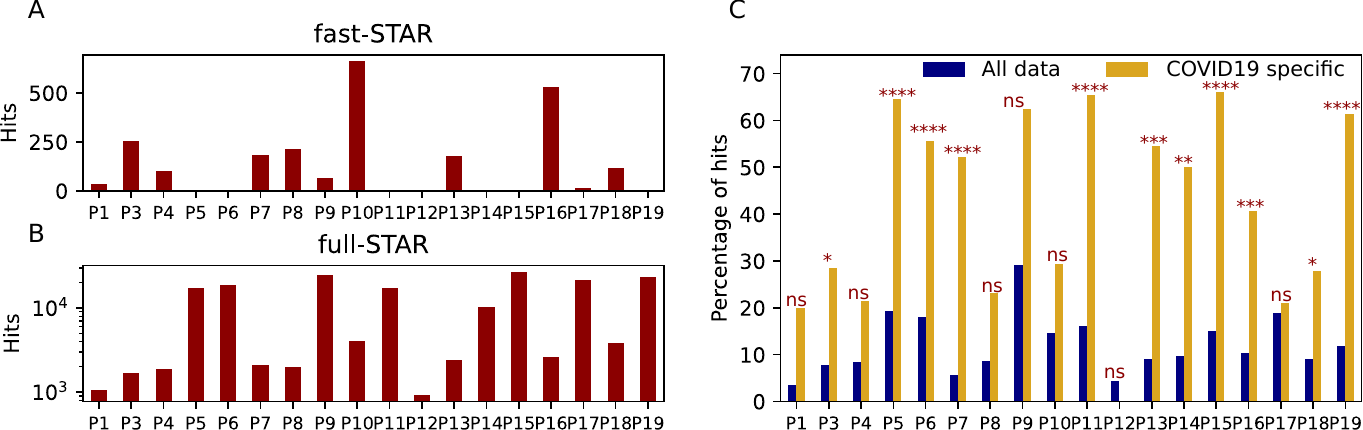}
    \caption{{\bf COVID19 specific sequences.} \textbf{A}: Number of hits obtained with fast-STAR per patient.
    \textbf{B}: Number of hits obtained with full-STAR per patient. \textbf{C}: Percentage of hits obtained with full-STAR in the entire dataset for each patient versus the percentage of hits obtained with full-STAR in the subsample of the dataset with Sars-CoV-2 specific sequences taken from the COVID19 antibody database, with significance. The non significant patients are labeled with \textbf{ns} ($p > 0.05$), one star corresponds to a pvalue $p \leq 0.05$, two stars $p \leq 0.01$, three stars $p \leq 0.001$ and four stars $p \leq 0.0001$.
      \label{Ab_database_covid}
      }
  \end{figure*}
  
\subsection*{Application to COVID-19 repertoires}

To assess the generalizability of our computational pipeline, we applied the same methodology to a distinct dataset from subjects naturally infected with COVID19. We used IgH repertoire data collected from Galson et al.~\cite{Osbourn}, comprising samples from 18 subjects at the infection peak. Note that no longitudinal data was available, providing a test case for our method. We applied the STAR pipelines
to identify candidate CDR3 amino acid heavy chain BCR sequences responding to COVID-19 (full list in {SI Table S2}). Applying fast-STAR we obtained an average of $131$ hits {(range 0 -- 661)}, while with full-STAR we obtained an average of $10,169$ {(range 918 -- 27,024)}. The clusters obtained with fast-STAR for each patient are shown in Fig. \ref{Covid_cluster}. In certain patients (10 out of 18), a polyclonal response was not detected, indicated by either zero or only one cluster. This observation may be attributed, in part, to variations in sequencing depth, as detailed in Table S2. Conversely, some patients exhibited an exceptionally diverse response, with up to 11 clusters observed in the case of patient 10. Notably, the highly conservative fast-STAR pipeline did not detect any shared hits between different patients. 

The source study did not test antibodies for specificity, making a direct validation of our candidate COVID-specific sequences difficult. Nonetheless, to assess the specificity of our method,
for each subject we evaluated the overlap between our hits and a comprehensive COVID19 database containing known antibodies associated to various variants of the SARS-CoV-2 virus \cite{raybould2021cov}.
For each subject, we compared the fraction of full-STAR hits within the full repertoire to the fraction of full-STAR hits among sequences from the repertoire that were also found in the COVID19 database. We found a significant enrichment of hits in the part of the repertoire that overlaped with the database in 11 out of the 17 subjects that we analyzed (Fig. \ref{Ab_database_covid}). This application demonstrates the versatility of our pipeline in identifying responsive BCR sequences, including in the context of a natural infection.

\section{Discussion}


The main interest of STAR is its ability to detect responding clonotypes from a single repertoire snapshot, without the need of longitudinal data. This method outperforms the naive approach of selecting clones with the highest frequencies as the likely responders.
We validated our results using a variety of tests, including binding assays and an analysis of overlap with previously published influenza datasets. 

Our approach is inspired by the ALICE method developed for T cells \cite{Alice}. Both methods share the idea of examining amino acid sequence neighbors as a measure of similarity,
and, in the case of full-STAR, to compare the number of neighbors to an expectation computed using a generative model of receptor sequences. 
The key difference arises from the nature of T cell responses, which do not form lineages. In T cells the emergence of clusters of neighboring sequences can only occur through the convergent selection for a common function of distinct lineages, while for B cells this effect is confounded by hypermutations, which create neighbors in sequence space belonging to the same lineage. Distinguishing the two effects is in general difficult, in particular because it is often hard to separate distinct lineages that have the same or similar CDR3 \cite{spisak2022combining,ralph2022inference,hoehn2022phylogenetic,nouri2020somatic,nouri2018spectral}. 

The STAR pipelines can in principle be extended to light-chain or paired-chain repertoire data. However, the low diversity of the light chain, both in terms of sequence length and variability, makes it less informative than the heavy chain. High-throughput paired-chain data (with at least 100,000 unique sequences) would ideally be the most informative, however most single-cell sequencing data are restricted to relatively small dataset (a few thousands), where the neighbors are only sparsely sampled, making the method inapplicable.
When both massive bulk single-chain repertoire ($>100,000$) and a smaller number (1,000-10,000) of single-cell data are available, hits identified by our method from the bulk data can be matched with paired chains from the single-cell data to infer the full antibody sequence, which can be subsequently tested for binding or neutralization.

Our findings highlight the importance of focusing on days close to the peak of infection, particularly day 7 in the case of influenza, for robust identification of responsive BCRs sequences. The algorithm's efficacy decays rapidly after the peak, underscoring the sensitivity of our method to the precise day when the sample is taken. Future investigations could explore the decay dynamics following the peak, and its implications for long-term immune responses.

While our pipeline exhibits promising results, certain limitations should be acknowledged. Fast-STAR is very specific, but likely misses a large fraction of responding clones. Full-STAR is more comprehensive, but may contain a substantial number of false positives, even if those contribute moderately to the cumulative frequency. The sensitivity of both methods strongly depends on sequencing depth, which must be sufficient to sample enough neibhbors of responding sequences. Further experimental tests should be applied to the candidate sequences proposed by STAR to properly assess their function.
The algorithm relies on a computational null model that assumes independence between B cells and ignores their possible lineage relationship. Because memory B cells have undergone hypermutations, we expect them to violate this assumption even in the absence of an immune challenge. In practice, this means that the null model should systematically underestimate the number of neighbors. Luckily, this effect is compensated by another inaccuracy. The approximate expected number of neighbors we use in full-STAR is actually an overestimation compared to the exact computation, because many amino acid neighbors are not proper antibody sequences and thus have extremely low probabilities, which is not accounted for in the approximation. As a result of these two errors canceling each other, the approximate estimate is in fact more accurate than the exact one. A multiplicative factor may be manually added to the exact estimate to correct for the error (Fig. \ref{Lambda_neigh}). However, future progress should rely on refining the null model to incorporate the generation of correlated lineages and thus provide a more accurate model of the human B cell repertoire before immunization. A more detailed treatment of the expected dynamics of B cell clones in absence of stimulation could also help improve the null model predictions.

We developped and validated our method in the context of immunization, and showed that it could also be applied to the case of a natural infection. Future research could test its applicability to other immune contexts, such auto-immune diseases \cite{Jiang2020}, allergies \cite{Levin2016}, or chronic infections such as HIV \cite{Johnson2018}. Compared to an acute challenge, the signal may be too weak to detect a response, unless repertoires are collected during a burst. Chronic disorders are expected to produce a higher number of highly mutated lineages \cite{Hoehn2021}, which may be more easily detected by the method.

The notion of convergent selection demonstrated here is related to that of public repertoires, defined as the set of receptors shared between individuals afflicted with the same condition \cite{Osbourn,Montague2021,ruiz2023modeling}. While many receptors are shared by chance due to their high generation probability, even in healthy individuals \cite{Elhanati2018,ruiz2023modeling}, individuals with a common condition tend to share more receptors, owing to the shared selective pressures that they are subjected to. Because of this shared convergent selection, one could expect the clusters of reactive BCRs identified by STAR to overlap between patients. On the contrary, we found no such sharing among COVID19 or influenza-vaccine patients. This implies that, while there exists a public repertoire responding to the same disease or vaccine, the primary immune response remains predominantly private. Understanding how to combine information from both private and public contributions to the response could help design better predictors of immune status.

\section{Methods}

We develop our method using data from mRNA--based heavy chain sequencing of the BCRs of 5 healthy human subjects vaccinated against influenza~\cite{Horns1}. The sequences were tagged with unique molecular identifiers (UMI) to correct for the PCR amplification bias. Pre-processed data was aligned to V, D and J templates with MiXCR \cite{MIXCR}, and then filtered to remove singletons.
Each repertoire contained on average $3.2\,\times 10^{5}$ {(range 15,035 -- 814,033)} unique UMI sequences, $6.0\,\times 10^{4}$  {(range 3,767 -- 121,608)} unique nucleotide sequences, and $5.2\,\times 10^{4}$  {(range 3,367 -- 101,254)} unique amino acid CDR3 sequences, refer to Table S1 for more details.

In each repertoire, for each amino acid heavy-chain CDR3 sequence we counted the number of its neighbors using the ATrieGC software~\cite{Natanael}, which uses indexing trees to efficiently span neighbors. Each neighbor is weighted by its multiplicity in terms of unique nucleotide sequences that have the corresponding amino acid CDR3.

The computational prediction for the number of neighbors (counted in terms of unique nucleotide sequences) is based on the OLGA software \cite{OLGA}, according to the formula:
\begin{equation}
\lambda(s)=N \sum_{s'\in \textit{V}(s)} P_{\rm gen}(s'),
\end{equation}
where $\lambda(s)$ is the expected number of neighbors of CDR3 amino acid sequence $s$, $V(s)$ is the set of neighbors of $s$ (one amino acid difference), $N$ the total number of unique nucleotide sequences of the repertoire, and $P_{\rm gen}(s')$ the generation probability of a CDR3 amino acid sequence $s'$ as given by the OLGA model.
Because of the high number of neighbors for each sequence, applying the formula above directly is computationally too expensive, but it may be estimated using a Monte-Carlo sample of $10^8$ sequences generated by OLGA, counting neighbors for each $s$, normalized by $10^8$, and re-multiplied by $N$.
In practice however, we used an approximation where we assume $P_{\rm gen}(s)\approx P_{\rm gen}(s')$,
 as justified by Fig. \ref{Pgen_Neigh}, which yields $\lambda(s)\approx 19 L(s) N P_{\rm gen}(s)$, where $L(s)$ is the CDR3 length of $s$. The results of the Monte-Carlo estimate of the exact formula, and of the approximate formula that we used in the paper, are compared with the pre-vaccination data in Fig.~\ref{Lambda_neigh}.

The full-STAR pipeline compares the expected number of neighbors $\lambda(s)$ with the observed one, denoted by $n(s)$. The p-value is computed as the probability to find at least as many neighbors as observed, where that number is assumed in the null model to follow a Poisson distribution of mean $\lambda(s)$, $p=\sum_{n=n(s)}^{\infty} \frac{e^{-\lambda(s)}}{n!}\lambda(s)^n$.

\section*{Data availibility}

The trivalent vaccine influenza bulk data
\cite{Horns1} and the single-cell data \cite{Horns2} are available on the European Nucleotide Archive with accession number, Sequence Read Archive: PRJNA512111, BioProject - PRJNA512111.\\ 
The vaccine influenza single-cell RNA sequencing and V(D)J data of \cite{Kleinstein} have been deposited in NCBI’s Gene Expression Omnibus and are available at the GEO Series accession number GSE175524.\\
The raw COVID-19 BCRs sequence data \cite{Osbourn} are available on the European Nucleotide Archive under BioProject Accession PRJNA638224.\\The sequences of the antibodies tested for being COVID-19 specific are available at \url{https://opig.stats.ox.ac.uk/webapps/covabdab/}.\\
The code to reproduce fast and full STAR is freely available at \url{https://github.com/statbiophys/STAR}.

\section*{Acknowledgements}
This work was supported by Sanofi, the European Research Council
consolidator grant no 724208 (AMW, TM, MFA), and the Agence Nationale
de la Recherche grant no ANR-19-CE45-0018 “RESP-REP” (AMW, TM,
MFA). EV and MAS are Sanofi employees and may hold shares and/or stock options in the company.
The authors declare that this study received funding from Sanofi. The funder collaborated directly in the study and was involved in the study design, analysis, and interpretation of data, the writing of this article, and the decision to submit it for publication.
The authors are grateful for the discussions and suggestions from Natanael Spisak, Emanuele Loffredo, Antoine Aragon and Maria Ruiz Ortega.

\bibliographystyle{pnas}

\onecolumngrid
\newpage
\twocolumngrid

\renewcommand{\thefigure}{S\arabic{figure}}
\setcounter{figure}{0}

\begin{figure} 
	\includegraphics[width=\linewidth]{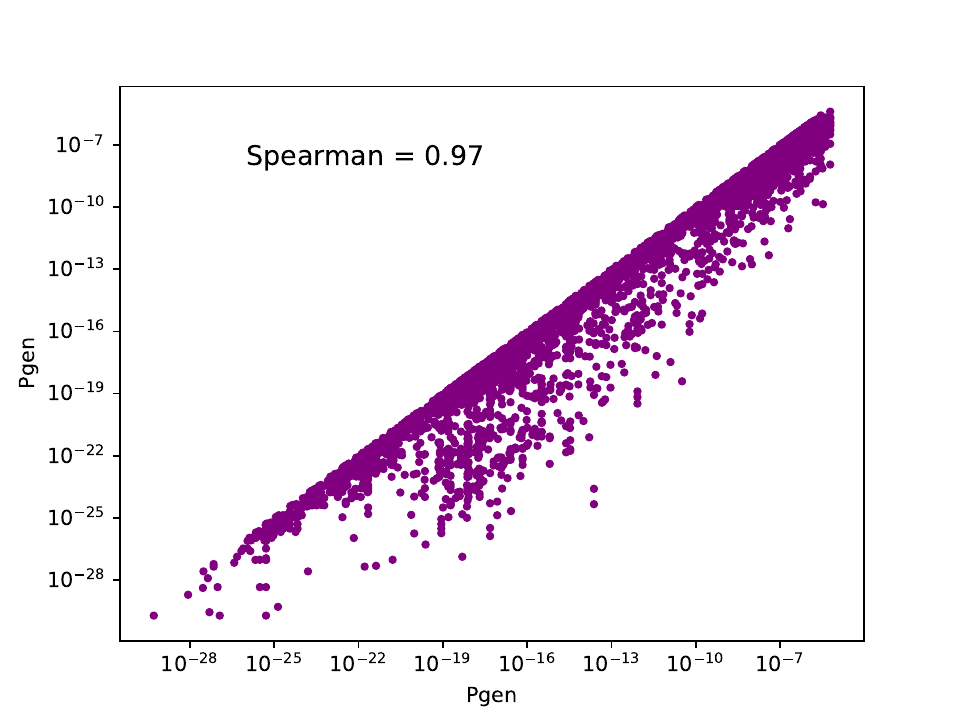}
        \caption{Comparison of generation probabilities of two neighboring CDR3 heavy-chain amino acid sequences, as computed by the OLGA software. Each point is a pair of neighboring sequences, with the one with the highest probability on the x axis.}
    \label{Pgen_Neigh}
  \end{figure}

\begin{figure*} 
    \includegraphics[width=\linewidth]{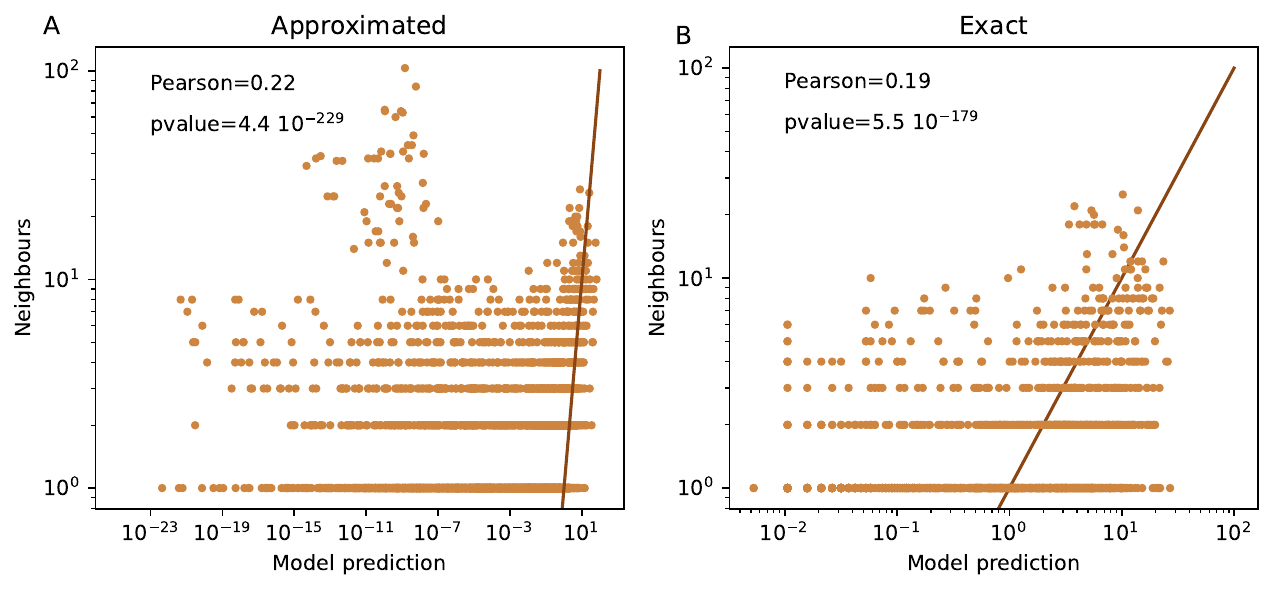}
    \caption{
Number of neighbors, data versus model prediction using {\bf (A)} the approximate estimate where the generation probability of all neighbors is assumed to be the same as the focal sequence, and {\bf (B)} the exact computation where the generation probabilities of all neighbors are summed. We computed the latter by simulating $10^8$ sequences using OLGA, and counting the number of neighbors.
Note that this exact computation underestimates the actual number of neighbors, requiring a corrective multiplicative factor (equivalent to a constant offset on the double-logarthmic scale shown here).
}
    \label{Lambda_neigh}
\end{figure*}

\begin{figure*} 
    \includegraphics[width=\linewidth]{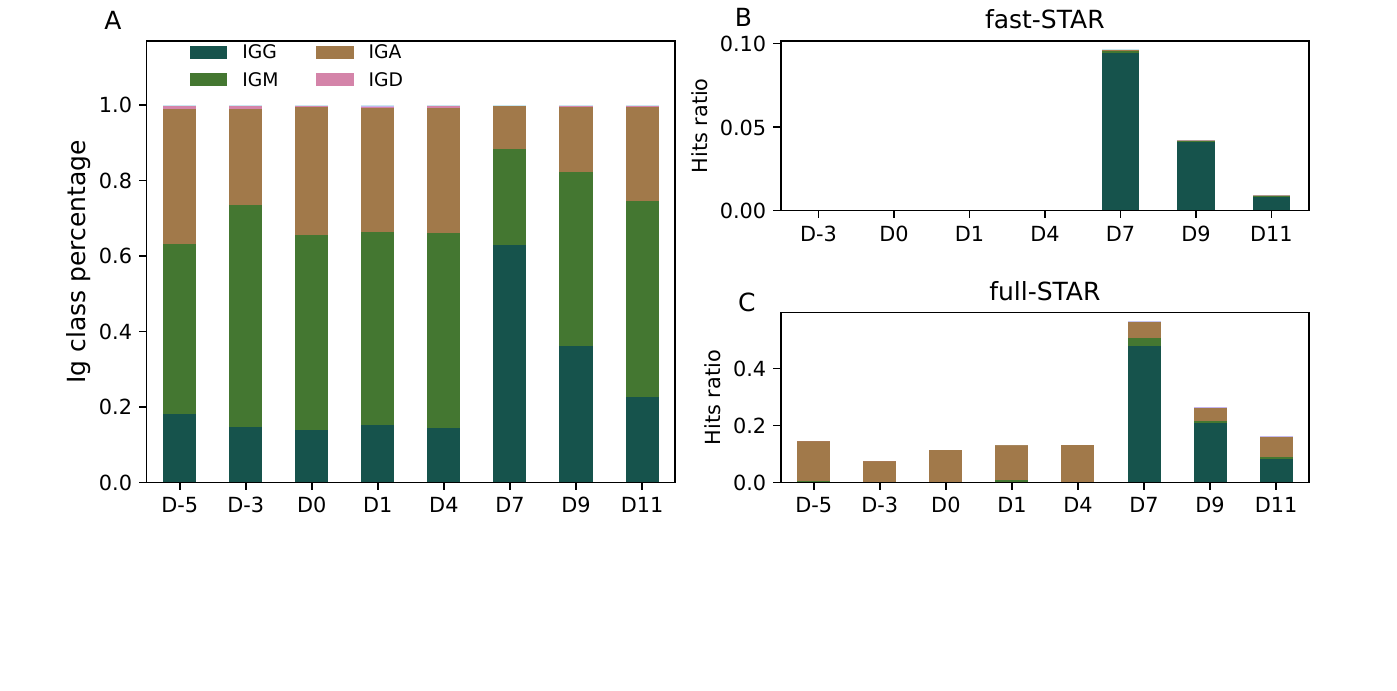}
    \caption{
{\bf A}. Distribution of immunoglobulin (Ig) classes as a function of days {\bf B.-C.} Cumulative frequency of hits by Ig class {\bf (B)} fast-STAR and {\bf (C)} full-STAR.
}
    \label{IgClass}
  \end{figure*}

\begin{figure*} 
    \centering
    \includegraphics[width=\linewidth]{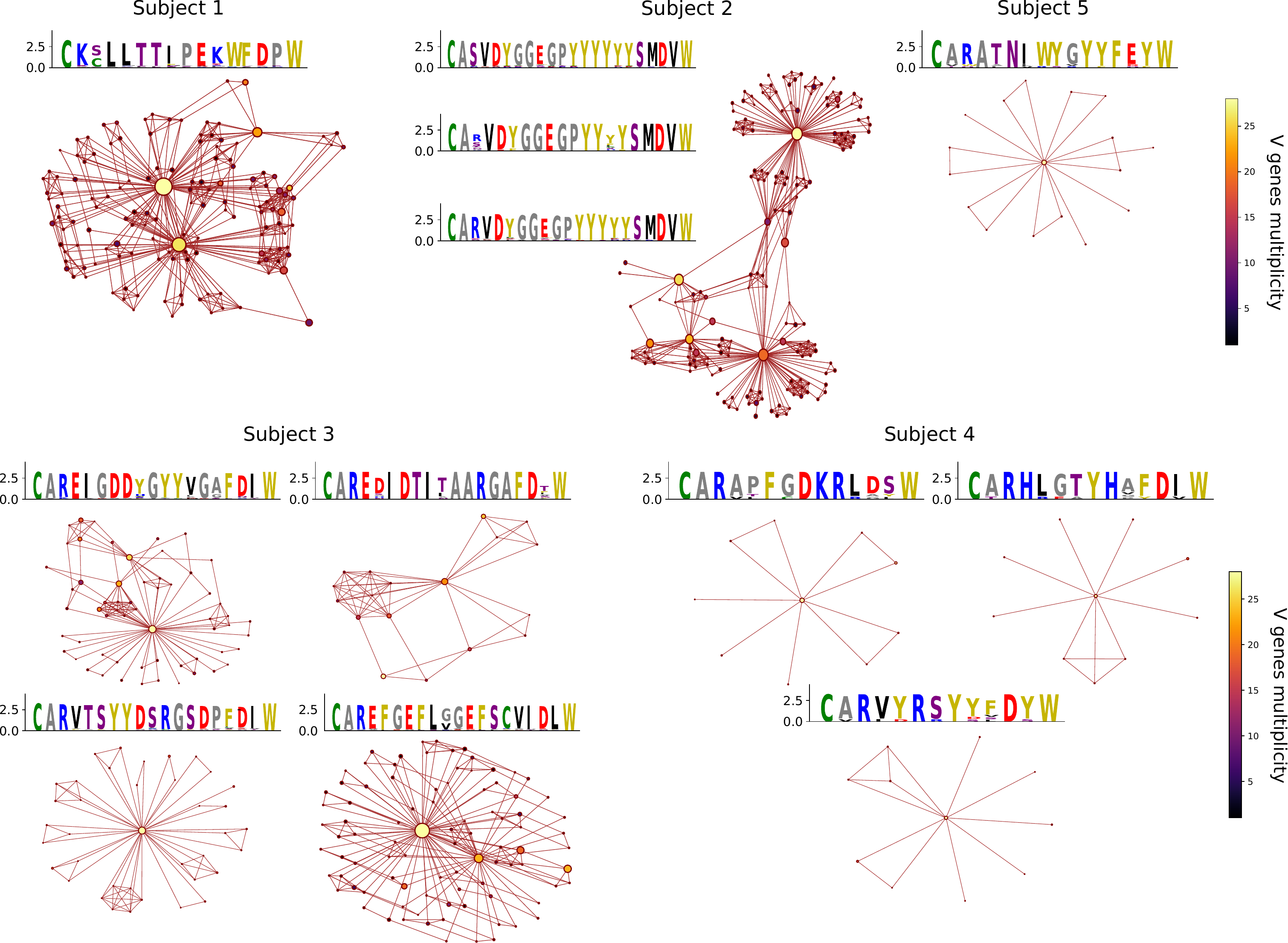}
    \caption{Graph structure of the responding antibodies for each subject, according to fast-STAR. An edge links two nodes (CDR3 amino acid sequences) if they differ by one amino acid. The size of each node is proportional to the number of unique nucleotides the CDR3 have (nucleotide multiplicity). Clusters with fewer than 10 sequences are not shown, and removed from the fast-STAR hits.}
    \label{clusterlogo}
  \end{figure*}

\begin{figure} 
    \includegraphics[width=\linewidth]{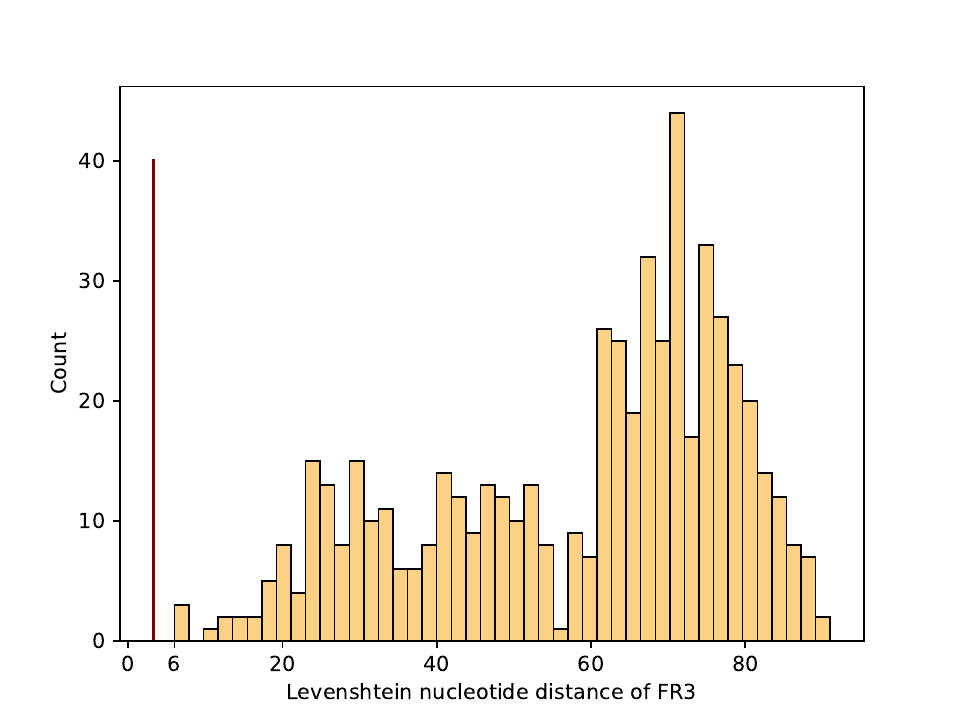}
    \caption{Distribution of Levenshtein distance between the nucleotide sequences of the germline V genes associated with the cluster of subject 1. The germline sequences is restricted to the part that was sequenced, corresponding roughly to the framework 3 region of length 111. The red line represents the average number of mutations over the same region.}
    \label{Vgene}
\end{figure}

\begin{figure*} 
    \includegraphics[width=\linewidth]
    {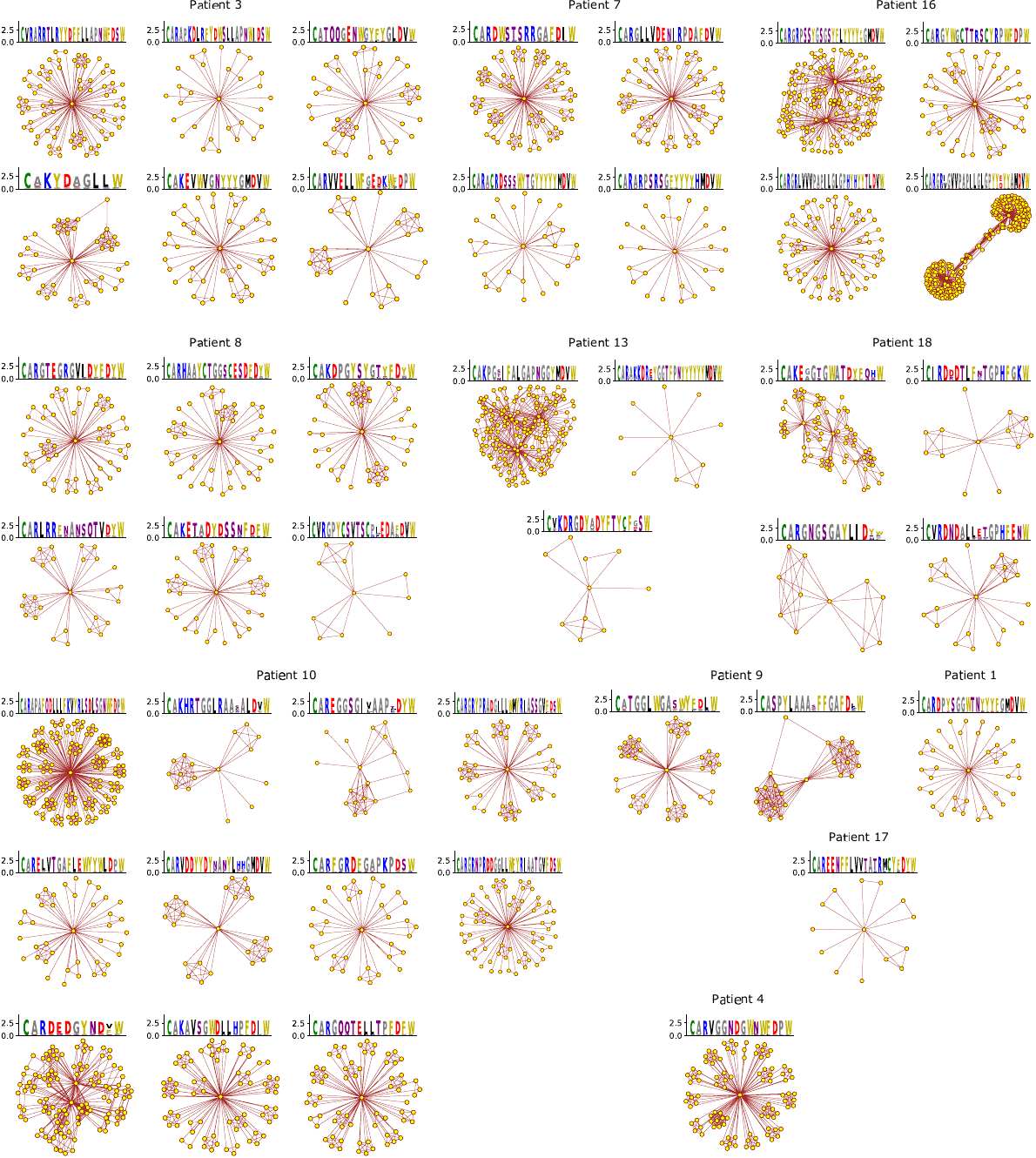}
    \caption{Graph structure of the responding antibodies for each patient with COVID-19, according to fast-STAR. An edge links two nodes (CDR3 amino acid sequences) if they differ by one amino acid. Clusters with fewer than 10 sequences are not shown, and removed from the fast-STAR hits.}
    \label{Covid_cluster}
\end{figure*}

\end{document}